\PassOptionsToPackage{unicode}{hyperref}
\PassOptionsToPackage{hyphens}{url}
\documentclass[11pt, ]{article}

\usepackage{mathtools}
\usepackage{amsmath}
\usepackage{amsthm}
\usepackage{amssymb}
\usepackage[italicdiff]{physics}
\mathtoolsset{showonlyrefs}

\usepackage{iftex}
\ifPDFTeX
  \usepackage[OT1,T1]{fontenc}
  \usepackage[utf8]{inputenc}
  \usepackage{textcomp} 
    \usepackage[p,osf,swashQ]{cochineal}
  \usepackage[cochineal,vvarbb]{newtxmath}
      \usepackage[scale=0.95]{biolinum}
    \usepackage[scale=0.95,varl]{inconsolata}
\else 
  \usepackage[scale=0.95,varl]{inconsolata}
  \usepackage{newpxtext}
  \usepackage{mathpazo}
    \usepackage[scale=0.95]{biolinum}
  \fi
\ifLuaTeX
  \usepackage{selnolig}  
\fi
\IfFileExists{microtype.sty}{
  \usepackage[]{microtype}
  \UseMicrotypeSet[protrusion]{basicmath} 
}{}

\allowdisplaybreaks
\usepackage[dvipsnames,svgnames,x11names]{xcolor}
\usepackage[lmargin=1in,rmargin=1in,tmargin=1in,bmargin=1in]{geometry}
\usepackage[format=plain,
  labelfont={bf,sf,small,singlespacing},
  textfont={sf,small,singlespacing},
  justification=justified,
  margin=0.25in]{caption}

\usepackage{sectsty}
\usepackage[compact]{titlesec}
\makeatletter
\newcommand\@shorttitle{}
\newcommand\shorttitle[1]{\renewcommand\@shorttitle{#1}}
\usepackage{fancyhdr}
\fancyheadoffset{0pt}
\makeatother
\renewenvironment{abstract}{ 
  \centerline
  {\large\sffamily\bfseries Abstract}\vspace{-1em}
  \begin{quote}\small
}{
  \end{quote}
}

\makeatletter
\newcommand{\assumplabel}[2]{%
   \protected@write \@auxout {}{\string\newlabel{#1}{{#2}{\thepage}{#2}{#1}{}}}%
   \hypertarget{#1}{#2}%
}
\makeatother

\usepackage{longtable,booktabs,array}
\usepackage{calc} 
\usepackage{etoolbox}
\makeatletter
\patchcmd\longtable{\par}{\if@noskipsec\mbox{}\fi\par}{}{}
\makeatother
\IfFileExists{footnotehyper.sty}{\usepackage{footnotehyper}}{\usepackage{footnote}}
\makesavenoteenv{longtable}
\usepackage{graphicx}
\makeatletter
\def\maxwidth{\ifdim\Gin@nat@width>\linewidth\linewidth\else\Gin@nat@width\fi}
\def\maxheight{\ifdim\Gin@nat@height>\textheight\textheight\else\Gin@nat@height\fi}
\makeatother
\setkeys{Gin}{width=\maxwidth,height=\maxheight,keepaspectratio}
\makeatletter
\def\fps@figure{htbp}
\makeatother

\theoremstyle{plain}

\usepackage[labelformat=simple]{subcaption}

\newcommand{\X}{\mathcal{X}}
\DeclareMathOperator{\proj}{proj}
\renewcommand{\bar}{\overline}

\newcommand{\R}{\ensuremath{\mathbb{R}}}

\newcommand{\N}{\ensuremath{\mathbb{N}}}

\DeclareMathOperator{\E}{\mathbb{E}}

\newcommand{\gvn}{\;\middle|\;} 

\makeatletter
\@ifpackageloaded{caption}{}{\usepackage{caption}}
\AtBeginDocument{%
\ifdefined\contentsname
  \renewcommand*\contentsname{Table of contents}
\else
  \newcommand\contentsname{Table of contents}
\fi
\ifdefined\listfigurename
  \renewcommand*\listfigurename{List of Figures}
\else
  \newcommand\listfigurename{List of Figures}
\fi
\ifdefined\listtablename
  \renewcommand*\listtablename{List of Tables}
\else
  \newcommand\listtablename{List of Tables}
\fi
\ifdefined\figurename
  \renewcommand*\figurename{Figure}
\else
  \newcommand\figurename{Figure}
\fi
\ifdefined\tablename
  \renewcommand*\tablename{Table}
\else
  \newcommand\tablename{Table}
\fi
}
\@ifpackageloaded{float}{}{\usepackage{float}}
\floatstyle{ruled}
\@ifundefined{c@chapter}{\newfloat{codelisting}{h}{lop}}{\newfloat{codelisting}{h}{lop}[chapter]}
\floatname{codelisting}{Listing}

\makeatother
\makeatletter
\@ifpackageloaded{caption}{}{\usepackage{caption}}
\@ifpackageloaded{subcaption}{}{\usepackage{subcaption}}
\makeatother
\makeatletter
\@ifpackageloaded{tcolorbox}{}{\usepackage[skins,breakable]{tcolorbox}}
\makeatother
\makeatletter
\@ifundefined{shadecolor}{\definecolor{shadecolor}{rgb}{.97, .97, .97}}
\makeatother

\usepackage[]{natbib}
\newenvironment{CSLReferences}[2]{ 
\bibliography{references.bib}
\clearpage
}{}

\usepackage{hyperref}
\usepackage{url}
\hypersetup{
  pdftitle={Projective Averages for Summarizing Redistricting Ensembles},
  pdfauthor={Cory McCartan},
  pdfkeywords={redistricting simulation, gerrymandering, partisan
dislocation},
  colorlinks=true,
  linkcolor={black},
  filecolor={Maroon},
  citecolor={VioletRed4},
  urlcolor={DodgerBlue4},
  pdfcreator={LaTeX via pandoc}}

\title{\sffamily\bfseries\huge\parfillskip=0pt
\rightskip=0pt plus .5\textwidth
\leftskip=0pt plus .5\textwidth
\emergencystretch=.3\textwidth Projective Averages for Summarizing
Redistricting Ensembles}
\shorttitle{Projective Averages for Summarizing Redistricting Ensembles}
\author{\textbf{Cory McCartan}\footnote{
To whom correspondence should be addressed.
Email: \texttt{\href{mailto:corymccartan@nyu.edu}{corymccartan@nyu.edu}}.
Website: \url{https://corymccartan.com/}.
Address:
60 5th Ave, New York, NY 10011.
The author thanks Christopher T. Kenny and Kosuke Imai for helpful
comments.}
\\Center for Data Science\\New York University
\vspace{0.05in}
 }
\date{January 11, 2024}

\begin{document}
\allsectionsfont{\sffamily}

\maketitle

\begin{abstract}
A recurring challenge in the application of redistricting simulation
algorithms lies in extracting useful summaries and comparisons from a
large ensemble of districting plans. Researchers often compute summary
statistics for each district in a plan, and then study their
distribution across the plans in the ensemble. This approach discards
rich geographic information that is inherent in districting plans. We
introduce the projective average, an operation that projects a
district-level summary statistic back to the underlying geography and
then averages this statistic across plans in the ensemble. Compared to
traditional district-level summaries, projective averages are a powerful
tool for geographically granular, sub-district analysis of districting
plans along a variety of dimensions. However, care must be taken to
account for variation within redistricting ensembles, to avoid
misleading conclusions. We propose and validate a multiple-testing
procedure to control the probability of incorrectly identifying outlier
plans or regions when using projective averages.
\end{abstract}

\textbf{\textit{Keywords}}\quad redistricting
simulation~\textbullet~gerrymandering~\textbullet~partisan dislocation
\ifdefined\Shaded\renewenvironment{Shaded}{\begin{tcolorbox}[enhanced, borderline west={3pt}{0pt}{shadecolor}, sharp corners, boxrule=0pt, breakable, interior hidden, frame hidden]}{\end{tcolorbox}}\fi


\hypertarget{sec-intro}{%
\section{Introduction}\label{sec-intro}}

Redistricting simulation algorithms are widely used in both academic
research on redistricting and in litigation over districting plans in
the United States \citep[see recently][]{imai2023sc}. Modern simulation
algorithms are able to generate a large sample, or \emph{ensemble}, of
districting plans from a specific target distribution that incorporates
real-world constraints on legislative districts \citep[most
recently][]{mccartan2023, autry2023metropolized, cannon2022spanning}.

Redistricting ensembles are commonly compared against a specific plan
(usually, the plan enacted by a legislature) in order to probe for
factors that affected its drawing. This comparison is done on the basis
of a number of district-level summary statistics. For instance, the
geographic compactness of each district in the enacted plan might be
compared to the corresponding measures for the ensemble. If the
statistics for the enacted plan are outliers in the distribution of the
statistic from the ensemble, that can be evidence of intentional
manipulation of the enacted plan away from the baseline represented by
the ensemble.

While analyses based on district-level statistics have been powerfully
applied to identify gerrymanders, the district-level approach leaves
something to be desired for the granular study of districting plans.
Redistricting is an intrinsically geographic endeavor, yet geography
makes no appearance in the district-level analysis: once the plans have
been simulated, we switch our focus to properties of abstractly numbered
districts, summarized with histograms and boxplots, not maps.
District-level analyses are also fundamentally coarse---if there are,
say, four districts, then the entire districting plan is boiled down to
just four numbers. It's impossible to use a district-level analysis to
conclude anything about the properties of a districting plan at much
more local scales.

This paper introduces the \emph{projective average} as a universal
operation for putting the geography back into district-level analyses.
The projective average operates on the fundamental building blocks of
districting plans: precincts, or in some states, counties. Informally,
the projective average takes a district-level summary statistic and
projects it back to the precinct level, producing a map that visualizes
the summary statistic for the simulation ensemble. It achieves this by
calculating the average value of that summary statistic in the various
districts to which the precinct belongs across the ensemble.
Section~\ref{sec-formal} formalizes this operation and demonstrates how
projective averages can be used to make comparisons between an ensemble
and a specific plan of interest.

Analyses using ideas similar to the projective average have appeared in
past work. Comparing a projective average to a particular plan (what we
term a \emph{projective contrast}) recalls the notion of \emph{partisan
dislocation} of \citet{deford2022partisan}, which can produce maps
suggesting where gerrymandering is occurring locally. But there are key
differences between the two techniques and their underlying goals.
Partisan dislocation was specifically designed as a ``simpler, far less
computationally intensive alternative to computer simulations,'' which
is calculated by averaging a variable number of precincts in a single
map. Moreover, partisan dislocation is designed to identify and quantify
U.S. partisan gerrymandering specifically, and as such applies only to
two-party vote shares (though it could in principle be extended to other
quantities). Projective averages, on the other hand, are a tool designed
for simulation ensembles specifically, and may be applied to any summary
statistic, such as compactness or racial composition. Thus projective
averages are a complementary technique to partisan dislocation measures,
rather than a replacement for them.

Despite these differences, other authors have referred to projective
contrast maps as ``dislocation maps'' \citep{imai2023sc}. These studies
and other similar work have all ignored the fundamental variation in
precinct-projected summary statistics across plans in the ensemble. As
explained in Section~\ref{sec-var}, some amount of discrepancy from the
ensemble-wide average is to be expected in any given plan. This natural
variation must be accounted for in order to properly contextualize
numerical results and avoid incorrectly concluding that a particular
part of a plan is an outlier.

\hypertarget{sec-formal}{%
\section{Formalizing Projective Averages}\label{sec-formal}}

\begin{figure}

\begin{minipage}[t]{0.30\linewidth}

{\centering 

\raisebox{-\height}{

\includegraphics{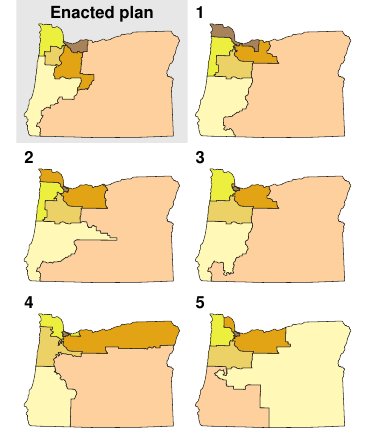}

}

}

\subcaption{\label{fig-ensemble}Enacted map and five sampled districting
plans from the simulation ensemble.}
\end{minipage}%
\begin{minipage}[t]{0.02\linewidth}

{\centering 

~

}

\end{minipage}%
\begin{minipage}[t]{0.30\linewidth}

{\centering 

\raisebox{-\height}{

\includegraphics{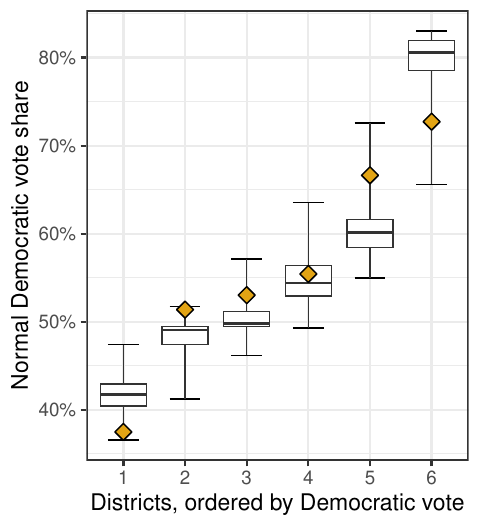}

}

}

\subcaption{\label{fig-distr-dshare}Distribution of Democratic vote
share order statistics for the ensemble. Values for the enacted plan are
marked by the blue points.}
\end{minipage}%
\begin{minipage}[t]{0.02\linewidth}

{\centering 

~

}

\end{minipage}%
\begin{minipage}[t]{0.36\linewidth}

{\centering 

\raisebox{-\height}{

\includegraphics{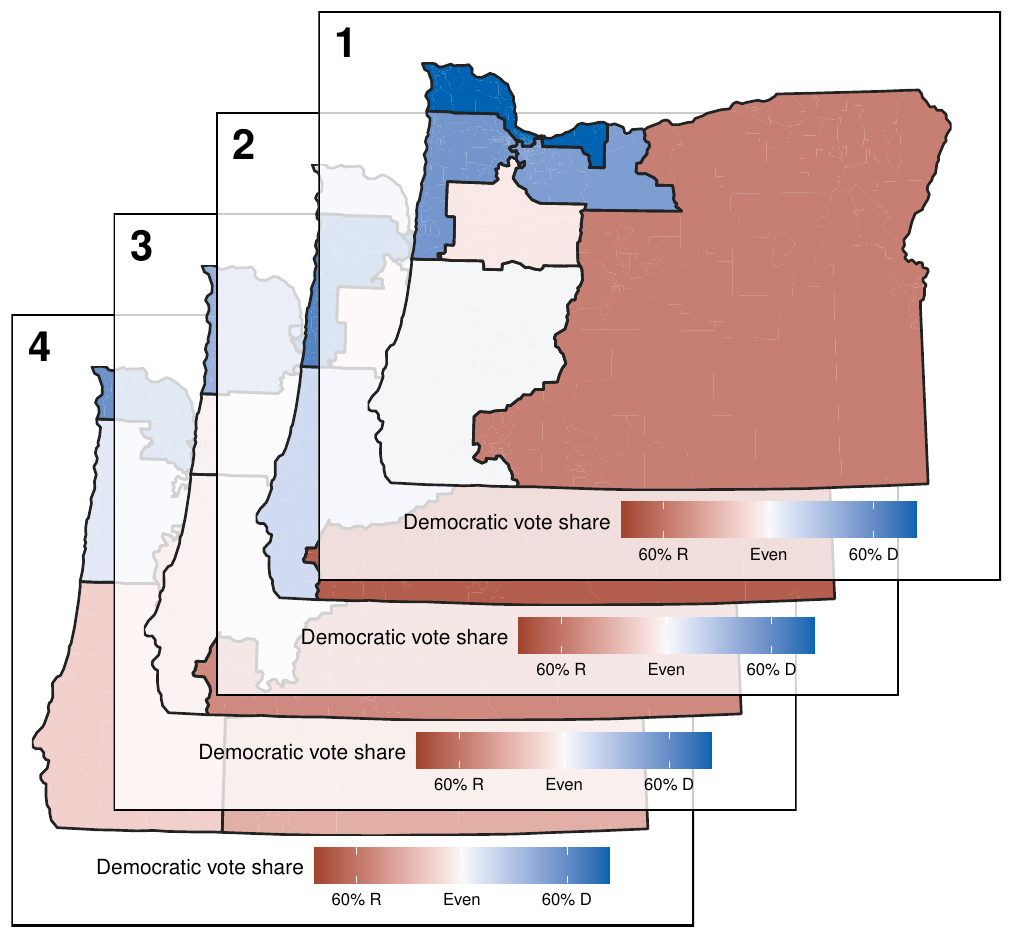}

}

}

\subcaption{\label{fig-proj-stacked}District-level Democratic vote
shares projected to the precinct level by plan.}
\end{minipage}%
\newline
\begin{minipage}[t]{0.48\linewidth}

{\centering 

\raisebox{-\height}{

\includegraphics{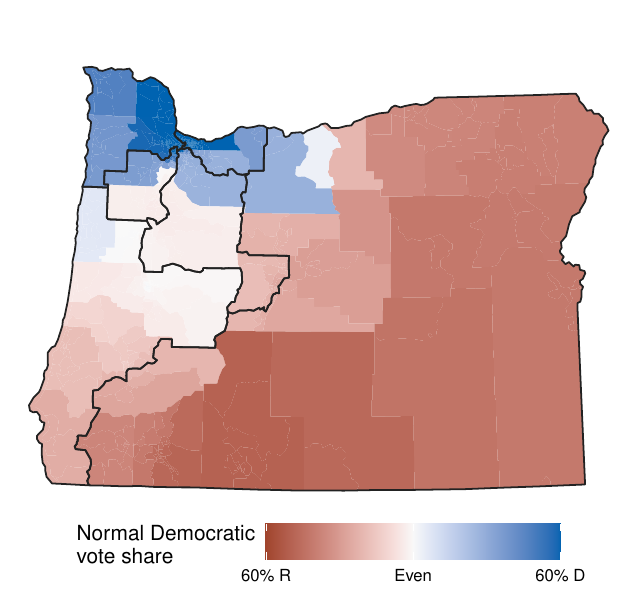}

}

}

\subcaption{\label{fig-proj-avg}The projective average of Democratic
vote share, with the enacted districts overlaid in black.}
\end{minipage}%
\begin{minipage}[t]{0.04\linewidth}

{\centering 

~

}

\end{minipage}%
\begin{minipage}[t]{0.48\linewidth}

{\centering 

\raisebox{-\height}{

\includegraphics{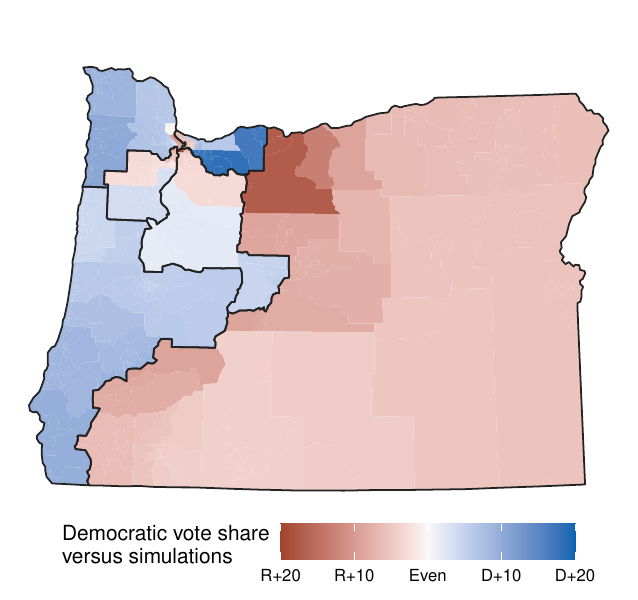}

}

}

\subcaption{\label{fig-proj-contr}The projective contrast of the enacted
and ensemble Democratic vote share. Blue indicates areas assigned to a
more Democratic district under the enacted plan than the average
simulated plan.}
\end{minipage}%

\caption{\label{fig-proj}Simulation analysis (a)--(b) and the process of
producing a projective average and contrast (c)--(e).}

\end{figure}

We motivate and demonstrate the use of projective averages in a
real-world context through a running example in the state of Oregon.
Between the 2010 and 2020 redistricting cycles, Oregon Democrats gained
unified control of state government. In 2021 the state adopted new
congressional districts on a party-line vote that some viewed as a
partisan gerrymander, though legal efforts to challenge the new map on
these grounds were unsuccessful \citep{opbcase}.

Simulation algorithms can be used to study the extent of partisan
gerrymandering in Oregon's 2021 districts by generating an ensemble of
partisan-neutral districting plans. In general, districting plans are
constructed out of precincts, counties, or other geographic ``building
blocks.'' We refer to the set of these geographic units (hereinafter
``precincts'' for simplicity) as \(V\). A districting plan with \(d\)
districts is then a mapping \(\xi:V \to \{1, 2, \dots, d\}\) from
precincts to integer labels. Districting plans must satisfy a number of
criteria, including equal population across districts, connectedness,
and geographic compactness. We denote the set of all valid districting
plans by \(\X\).

A district-level summary statistic \(f:\X\to\R^d\) reports a value for
each district in a plan. Generally \(f\) is defined by
\(f(\xi)_j = g(\xi^{-1}(j))\) for some function \(g:2^V\to\R\) that
calculates a summary value for any set of precincts. For instance, \(g\)
might calculate the typical Democratic vote share in a set of precincts.

A redistricting ensemble is a sample \(S=\{\xi_i\}_{i=1}^n\) of size
\(n\) from some probability distribution \(\pi\) on \(\X\). Here, we'll
use an ensemble of \(n=5000\) plans generated by
\citet{mccartan2022simulated} using the algorithm of
\citet{mccartan2023}. The target distribution for this sample
prioritizes geographically compact districts among those plans which are
contiguous and have all district populations within 0.5\% of equality.
Five sampled plans and the enacted districts are shown in
Figure~\ref{fig-ensemble}.

The Oregon ensemble's distribution of the order statistics of the normal
Democratic vote share are shown in Figure~\ref{fig-distr-dshare}, with
the enacted plan's values marked for comparison. For most districts, the
enacted plan's Democratic vote share lies in the tails of the simulated
distribution, with an overall pattern that suggests the packing of
Republican voters into one district and the cracking of Democratic
voters in the most heavily Democratic district. However, since
Figure~\ref{fig-distr-dshare} shows only order statistics, we are unable
to localize any of these patterns to a sub-district scale.

\hypertarget{projective-distribution-and-average}{%
\subsection{Projective distribution and
average}\label{projective-distribution-and-average}}

The projective average is designed to tackle these issues. We define the
\emph{projective distribution of} \(f\) \emph{at precinct} \(v\in V\) as
the distribution of \(f\) in the district which \(v\) belongs to, i.e.,
\[\proj_S f(v) \coloneq \{ f(\xi_i)_{\xi_i(v)} \}_{i=1}^n.\]

The \emph{projective average of} \(f\) \emph{at precinct} \(v\) is then
naturally defined as \[
\bar{f}_S(v) \coloneq \frac{1}{n} \sum_{i=1}^n f(\xi_i)_{\xi_i(v)}.
\] Thus while the district-level distribution of \(f\) is supported on
\(\R^n\), the projective distribution \(\proj_S f\) is supported on
\(\R^{|V|}\), and the projective average \(\bar f\) assigns each
precinct a single value, allowing it to be plotted on a map.

Figure~\ref{fig-proj-stacked} shows a sample of four elements from the
projective distribution of normal Democratic vote share, and
Figure~\ref{fig-proj-avg} plots the projective average, which is the
result of averaging across the maps in Figure~\ref{fig-proj-stacked} and
the 4,996 others in the ensemble. Because of the need to group voters
into districts, the projective average appears to ``smooth out'' the
partisan geography of the state

\hypertarget{projective-contrasts}{%
\subsection{Projective contrasts}\label{projective-contrasts}}

The same projection operation applied to an ensemble to yield the
projective distribution and average can be applied to a single
redistricting plan. Here, the operation simply maps each precinct to the
value of the summary statistic in its corresponding district. Thus, for
the enacted plan \(\xi^0\), we can identify \(\proj_{\{\xi^0\}} f\) with
\(\bar{f}_{\{\xi^0\}}\), each of takes on at most \(d\) values.

A traditional district-level analysis compares \(f(\xi^0)\) with
\(f(S)\). The analogue with projective averages is the \emph{projective
contrast}, formally defined as \(\bar{f}_{\{\xi^0\}} - \bar{f}_S\).
Intuitively, it reports how much higher or lower \(f\) is in the enacted
plan than the average simulated plan. For example, if \(f\) measures
districts' compactness, then the projective contrast would be positive
in regions of the state that belong to more compact districts under the
enacted plan than under the typical plan from the ensemble.

Figure~\ref{fig-proj-contr} plots the projective contrast of normal
Democratic vote share. Blue areas have positive values of the contrast
and correspond to precincts belonging to a more Democratic district in
the enacted plan than the average simulated plan. While small on the
statewide map, the city of Portland is red, indicating it is assigned to
a less Democratic district under the enacted map, while the areas
surrounding Portland and extending down the coast are blue. Meanwhile,
the south and east of the state are more Republican than would otherwise
be expected. This overall pattern is consistent with the ``crack heavily
Democratic areas, pack Republican areas'' strategy inferred from the
boxplots, but the projective contrast map allows us to confirm the
pattern and see exactly where the packing and cracking are occurring.

Projective averages and contrasts can also be re-aggregated to larger
geographies as a way of further summarizing the granular results of
projective averaging. This technique is formalized and demonstrated in
Appendix~\ref{sec-app-agg}. Population-level analogues of all of these
definitions are in Appendix~\ref{sec-app-pop}.

\hypertarget{sec-var}{%
\section{Accounting for Variation Across the Ensemble}\label{sec-var}}

Calculating a projective average requires compressing the entire
projective distribution over each precinct to a single number: its mean.
One side effect of this averaging step is to lose information about how
unusual a particular value for the projective average in a precinct is.
For instance, a plan that is more or less Democratic than the average
simulated plan may not be a partisan gerrymander or even an outlier in
the projective distribution. Informally, we would actually \emph{expect
the typical map to be atypical} in some areas of the state.

\begin{figure}

{\centering \includegraphics{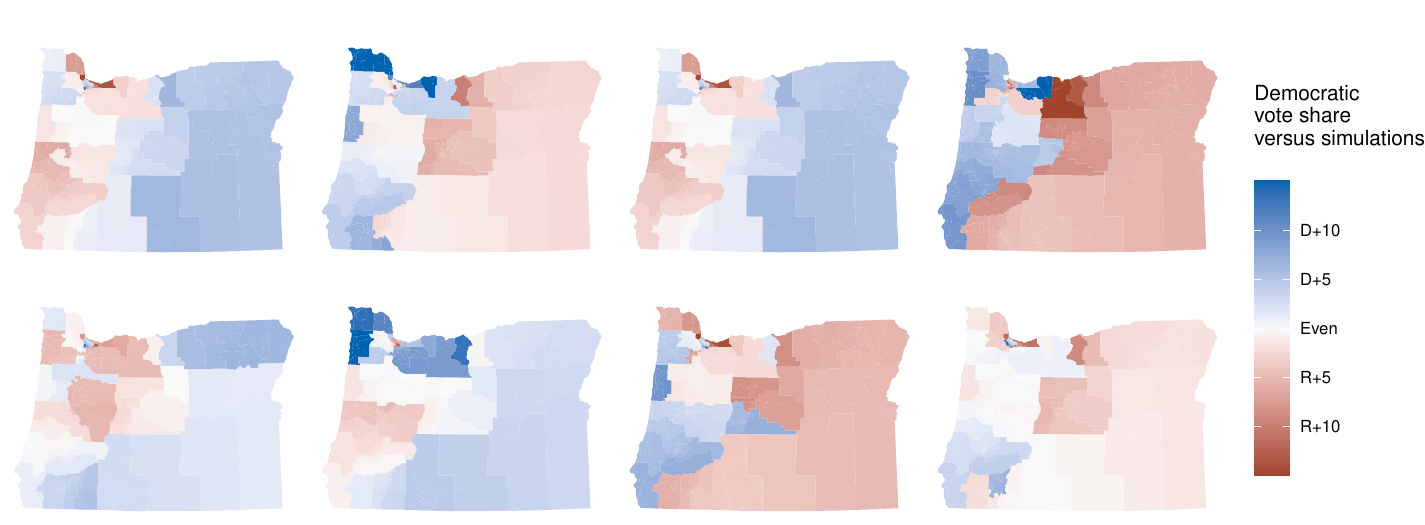}

}

\caption{\label{fig-contr-samp}Projective contrasts of seven randomly
sampled districting plans and the enacted plan versus the simulated
ensemble.}

\end{figure}

Figure~\ref{fig-contr-samp} illustrates the problem. The projective
contrast of Democratic vote share has been calculated, not just for the
enacted plan, but for seven draws from the ensemble as well. We invite
the reader to guess which of the eight plans is in fact the
enacted.\footnote{\rotatebox[origin=c]{180}{The enacted plan is in the upper right corner.}}
The difficulty of this task illustrates that the enacted plan may not be
an outlier along the dimensions plotted by the projective contrast. Put
differently, without accounting for the expected variation in projective
averages and contrasts across the ensemble, analysts may incorrectly
claim that a plan or district is gerrymandered, or unusually noncompact,
or so on.\footnote{Of course, if an omnibus statistical test has already
  been performed to establish that the enacted map is an outlier, the
  approaches advanced here may not be necessary. Projective contrasts
  and averages can illustrate or explain features of the plan, showing
  the ``how'' of a gerrymander whose existence has been rigorously
  established by other means. But the natural tendency to notice unusual
  patterns in a map may lead to informal conclusions about ``outlier
  regions'' of the map that would be better made in consultation with
  the techniques outlined here.}

Informal graphical approaches for addressing this problem are discussed
in Appendix~\ref{sec-app-graphic}. Here, we propose a more formal way of
testing when a projective contrast indicates deviations beyond what
would be expected for a random sample from the target distribution,
using multiple testing methodology from \citet[hereinafter
ST]{storey2001estimating}. The ST method is designed to estimate and
thereby control the \emph{positive False Discovery Rate} (pFDR), defined
as \[
\mathrm{pFDR} \coloneq \E\left[\frac{\text{false discoveries}}{\text{total discoveries}}\gvn \text{total discoveries} > 0\right],
\] for a collection of valid but possibly dependent p-values. We can
obtain a valid p-value for each precinct in the map by comparing the
projected value for the enacted map to the projective distribution at
that precinct.

\begin{figure}

{\centering \includegraphics{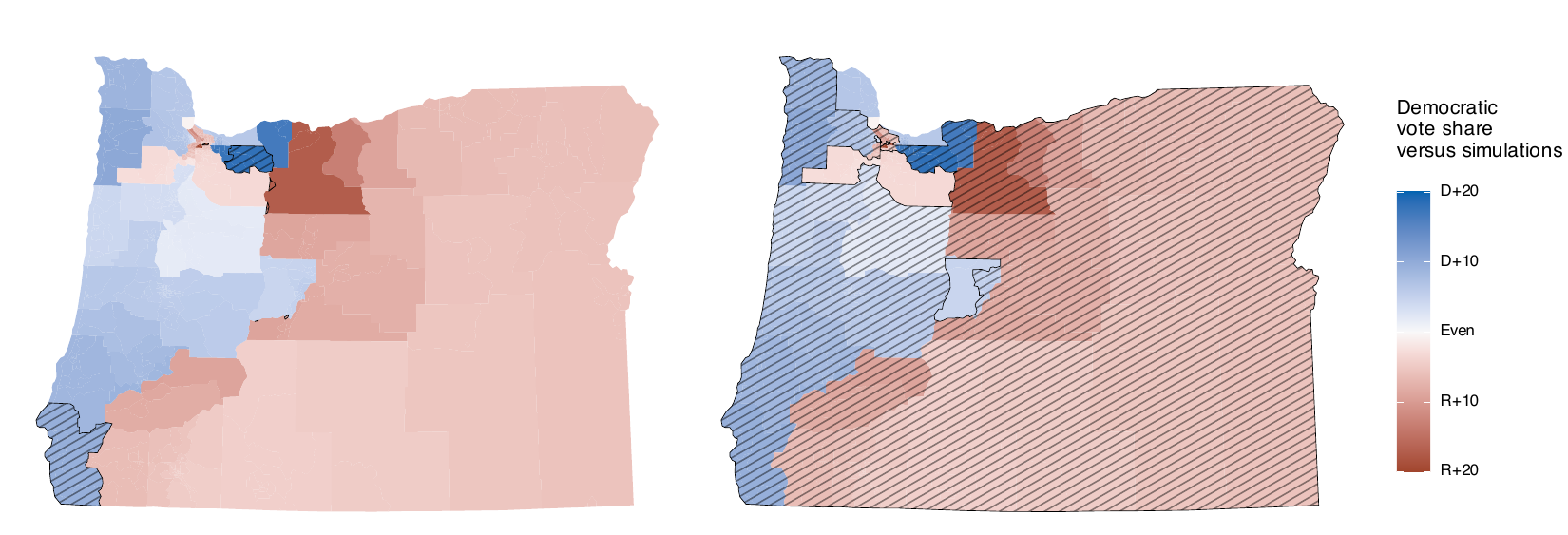}

}

\caption{\label{fig-fdr-ctrl}Projective contrast of the enacted and
ensemble Democratic vote share. Significant discrepancies from the
ensemble after controlling the pFDR at 1\% (left) and 5\% (right) are
marked by the hatched areas.}

\end{figure}

The ST methodology can be used to define a procedure, detailed in
Appendix~\ref{sec-app-st}, that takes the set of p-values for all the
precincts and returns a subset of ``discovered'' precincts for which the
enacted map can be considered to differ significantly from the ensemble.
Control of the pFDR at (say) the 10\% level means that, informally, at
most 10\% of the selected precincts can be expected to in fact not be
true departures from the ensemble's projective distribution (false
discoveries). Compared to more familiar FDR control approaches like the
Benjamini-Hochberg (BH) procedure, the ST procedure handles arbitrary
dependence while maintaining high statistical power. As
Appendix~\ref{sec-app-fwer} demonstrates, in the redistricting setting
the ST procedure is able to control the pFDR at a range of levels under
the global null distribution without being overly conservative.

Figure~\ref{fig-fdr-ctrl} demonstrates the ST methodology for two
different pFDR control levels, with the selected precincts hatched.
Almost all of the state is selected at the 5\% level, indicating that
the observed discrepancies measured by the projective contrast go beyond
what would be reasonably expected by random variation in the null
distribution. In contrast, \emph{none} of the simulated plans in
Figure~\ref{fig-contr-samp} are selected by the ST procedure, indicating
that those discrepancies, while visually of similar magnitude to the
enacted plan, are in fact within the expected range.

\hypertarget{sec-concl}{%
\section{Conclusion}\label{sec-concl}}

We have introduced the projective average and contrast as a technique
for visualizing and understanding district-level summary statistics
calculated from a redistricting simulation. While the illustrative
example here used Democratic vote share, we stress that projective
averages are useful far beyond studies of partisan gerrymandering, and
can be applied to any district-level summary, including compactness
scores or win probabilities. Appendix~\ref{sec-app-fdr-ex} contains
demonstrations for some of these other quantities. Re-aggregating
projective averages to predefined sub-state geographies, as demonstrated
in Appendix~\ref{sec-app-agg}, is also particularly useful.

No matter how the projective operation is used, it is important to
consider sampling variability and use caution before drawing statistical
conclusions---just as when working with district-level summary
statistics. We have proposed a statistical method for doing so, which we
recommend for routine work with projective averages and contrasts.

\hypertarget{refs}{}

\begin{CSLReferences}{0}{0}\end{CSLReferences}

\appendix

\hypertarget{sec-app-pop}{%
\section{Population projective averages and
contrasts}\label{sec-app-pop}}

Projective averages and contrasts can be defined for the overall
population distribution \(\pi\) of districting plans rather than for a
finite ensemble \(S\).

Letting \(\Xi\sim\pi\) be a single randomly sampled plan, \(f(\Xi)\) is
the distribution of the summary statistic in the \(n\) districts. If
\(\xi^0\) is the comparison plan of interest, the traditional
district-level analysis compares \(f(\xi^0)\) to the distribution
\(f(\Xi)\). As noted in the main text, in practice, analysts have access
not to the full distribution \(\pi\) but a finite ensemble
\(S=\{\xi_i\}_{i=1}^n\) for some \(n\).

We define the population projective distribution of \(f\) at precinct
\(v\in V\) as \[\proj_\pi f(v) \coloneq f(\Xi)_{\Xi(v)}.\] In other
words, the distribution of \(f(\Xi)\) in the (random) district which
\(v\) belongs to. When \(f\) is defined by a precinct-set summary
function \(g\), the projective distribution is equivalently defined by
\(\proj_\pi f = g\circ \Xi^{-1} \circ \Xi\), i.e., \(g\) applied to the
fiber of \(\Xi\) containing \(v\). The population projective average of
\(f\) at precinct \(v\) is then naturally defined as
\[\bar{f}_\pi(v) \coloneq \E[\proj_\pi f(v)] = \E[f(\Xi)_{\Xi(v)}],\]
and the projective contrast as \(\bar{f}_{\{\xi^0\}} - \bar{f}_\pi(v)\).

\hypertarget{sec-app-agg}{%
\section{Aggregating projective averages and
contrasts}\label{sec-app-agg}}

Projective averages and contrasts can also be re-aggregated to larger
geographies as a way of further summarizing the very granular results of
projective averaging. Aggregation is accomplished by taking a population
or voting-population weighted average of the projective average or
contrast across precincts. Formally, given a function \(q:V\to\N\) that
labels regions of the map, we can define an \emph{aggregated projective
average} as \[
\bar{F}^q_S(j) \coloneq \frac{\sum_{v\in q^{-1}(j)} w_v \bar{f}_S(v)}{\sum_{v\in q^{-1}(j)} w_v}
\] for each \(j\) in the range of \(q\) and for some population weights
\(\{w_v\}_{v\in V}\). Analogous definitions are naturally made for
population projective averages \(\bar{f}_\pi\) and for projective
contrasts such as \(\bar{f}_{\{\xi^0\}} - \bar{f}_S\).

\citet[Figure 2]{kenny2023widespread} take projective contrasts of the
Democratic win probability and then aggregate these contrasts up to the
congressional district level to form a national accounting of partisan
effects in each district. The use of aggregated projective contrasts
means that effects can be well-defined and calculated for each enacted
district, even though the underlying simulation ensembles involve
thousands of completely different, non-overlapping districts.

\begin{figure}

{\centering \includegraphics{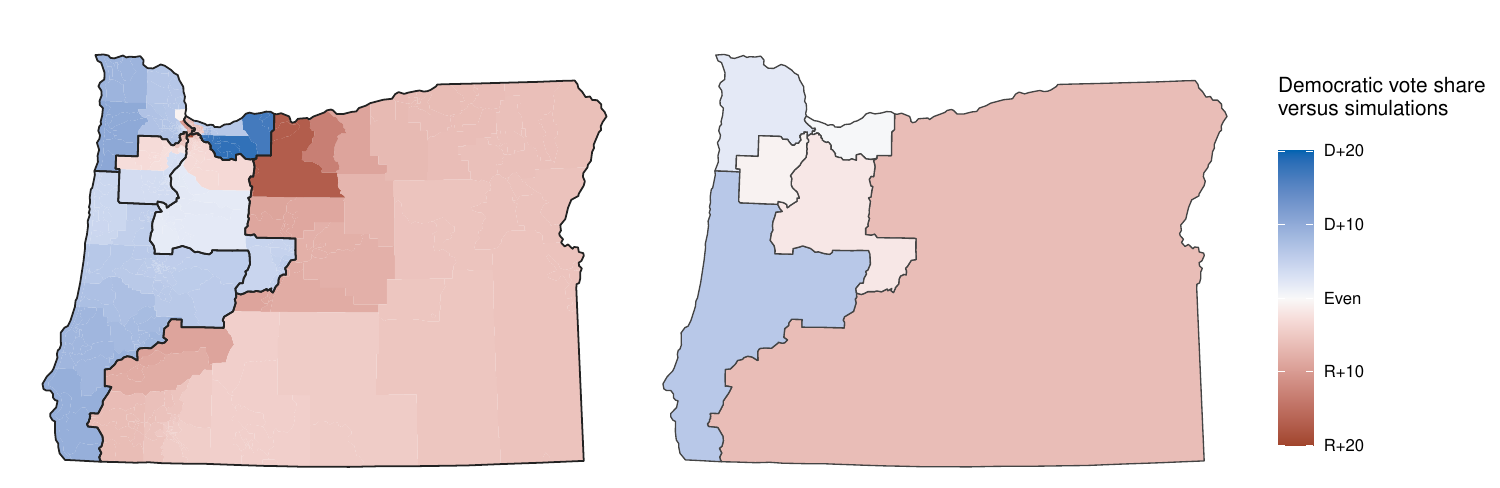}

}

\caption{\label{fig-proj-agg}Precinct-level projective contrast for
normal Democratic vote share (left), and aggregated to the enacted
districts (right).}

\end{figure}

We demonstrate this same idea for the Oregon ensemble in
Figure~\ref{fig-proj-agg}. Voters in blue districts on average belong to
a more Democratic district than would otherwise be expected from the
ensemble. Thus district 4 in the southwest appears to have been
gerrymandered to be more Democratic, while district 2 in the east
appears to have been gerrymandered to be more Republican.

\hypertarget{sec-app-st}{%
\section{Storey-Tibshirani pFDR control procedure for projective
contrasts}\label{sec-app-st}}

We combine Algorithm 1 of \citet{storey2001estimating}, which can be
applied to any set of dependent p-values, with a final search step to
identify the rejection region that is expected to control the pFDR at
the requested level. Storey and Tibshirani justify the procedure
theoretically (i.e., that it controls the FDR in the large-\(|V|\)
limit) in their article for ``clumpy'' dependence that would be expected
in redistricting problems. Empirical validation of this approach is
provided in Appendix~\ref{sec-app-fwer} for the Oregon ensemble. The
algorithm with our choice of its tuning parameter is detailed below for
completeness.

Let \[
p_v(\xi) \coloneq \frac{1+\#\{\xi'\in S:f(\xi')_{\xi'(v)} \ge f(\xi)_{\xi(v)}\}}{n}
\] be the p-value for the plan \(\xi\) in the projective distribution
\(\proj_S f(v)\) for a precinct \(v\in V\), where \(\#\{\cdot\}\)
indicates the cardinality of a set. Then we can calculate the observed
p-value \(p_v(\xi^0)\) for the enacted plan as well as \(n\) null
p-values \(\{p_v(\xi_i)\}_{i=1}^n\) for each precinct.

For a threshold \(\gamma\in[0,1]\), let
\(R(\gamma)\coloneq \#\{v\in V: p_v(\xi^0)\le\gamma\}\) be the number of
(actual) precinct p-values at or below the threshold. Define the
following average number of null rejections for a given threshold
\(\gamma\) by \[
\bar R_0(\gamma) \coloneq \frac{1}{n}\sum_{i=1}^n \#\{v\in V: p_v(\xi_i)\le\gamma\}.
\]

Then estimate the fraction of false discoveries as \[
\widehat\pi_0 = \frac{|V| - R(0.5)}{|V| - {\bar R_0}(0.5)}.
\]

For any \(\gamma\in[0,1]\), estimate the pFDR for thresholding the
p-values at \(\gamma\) by \[
\widehat{\mathrm{pFDR}}(\gamma) = \frac{\widehat\pi_0 {\bar R_0}(\gamma)}{R(\gamma)}.
\] To control the pFDR at a chosen level \(\alpha\in[0,1]\), select all
p-values at or below \[
\hat\gamma \coloneq \max\{\gamma\in[0,1] : \widehat{\mathrm{pFDR}}(\gamma)\le \alpha\}.
\] In practice, this may be achieved by calculating
\(\widehat{\mathrm{pFDR}}(\gamma)\) across a few dozen \(\gamma\) and
then interpolating linearly or with a monotone cubic spline. The
replication code takes this approach.

\hypertarget{sec-app-fwer}{%
\section{Empirical performance of pFDR control method in the null
distribution}\label{sec-app-fwer}}

\begin{figure}

{\centering \includegraphics{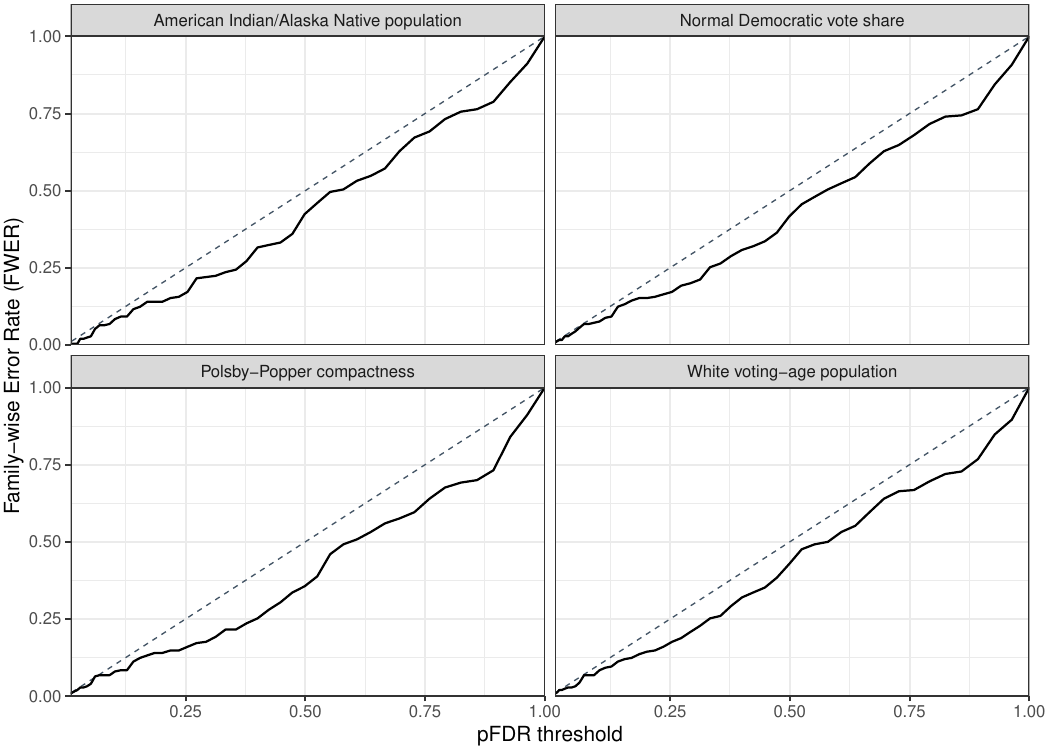}

}

\caption{\label{fig-fwer-null}Family-wise error rate across a range of
nominal pFDR control levels for the procedure described in
Section~\ref{sec-var} applied to the global null distribution from the
simulated ensemble.}

\end{figure}

When the global null hypothesis (that the enacted map was generated from
the same distribution as the ensemble) is true, pFDR control also
implies control of the \emph{family-wise error rate} (FWER), the
probability that any of the discoveries is false. FWER control can be
checked by applying the procedure to samples from the null
distribution---the simulated ensemble. This is done in
Figure~\ref{fig-fwer-null} for a range of nominal pFDR/FWER thresholds
and for four different test statistics. At no point does the actual FWER
exceed the nominal pFDR threshold. Importantly, however, the procedure
is also not overly conservative, with the observed FWER falling just
below the nominal pFDR level.

\hypertarget{sec-app-graphic}{%
\section{Graphical methods for accounting for variation in the
projective distribution}\label{sec-app-graphic}}

One simple, if not particularly formalized, way to address the problem
of variation in projective distributions is by using something exactly
like Figure~\ref{fig-contr-samp}, where the `true' projective contrast
or average is displayed among a number of `null' averages that provide
appropriate comparative context. Indeed, following the ``lineup
procedure'' of \citet{buja2009statistical} analysts may benefit from
initially randomizing the location of the true contrast and attempting
to identify it as a visual outlier.

\begin{figure}

{\centering \includegraphics{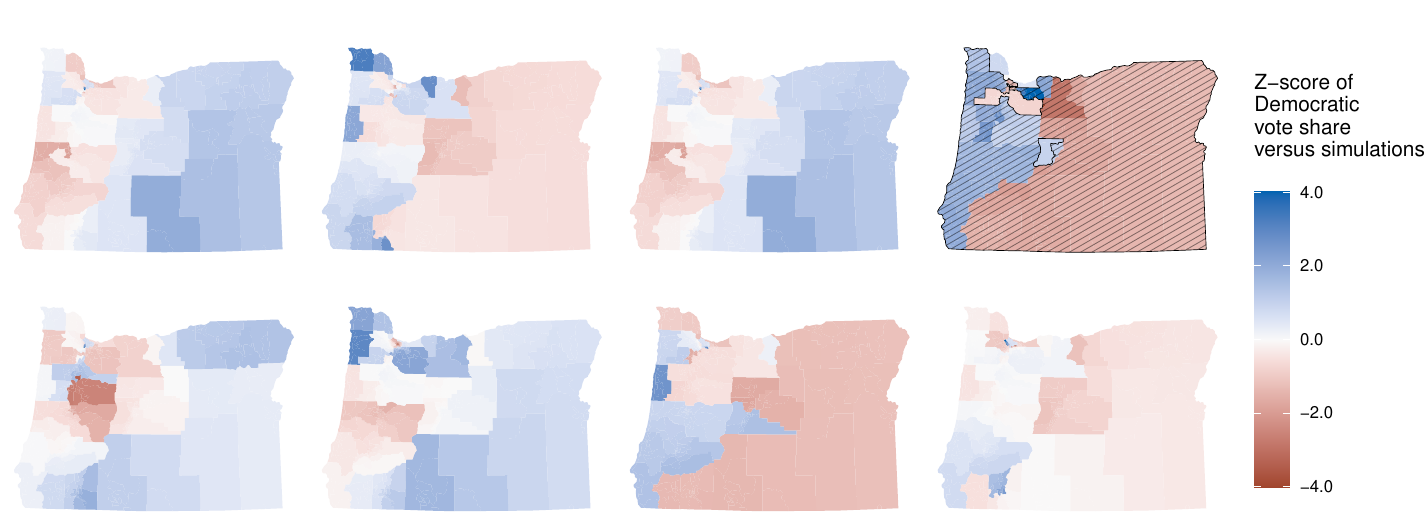}

}

\caption{\label{fig-samp-norm}Normalized projective contrasts of seven
randomly sampled districting plans and the enacted plan versus the
simulated ensemble. The pFDR control procedure has been applied to all
eight plans, with selected areas hatched.}

\end{figure}

A complementary approach is to normalize the projective contrast
precinct-wise by the standard deviation of the projective distribution
at each precinct. This has the effect of generating precinct-level
z-scores for the contrast of interest. Areas with relatively high
variation in the projective distribution will stand out less than before
normalization. A version of Figure~\ref{fig-contr-samp} with the
projective contrast normalized in this way is shown in
Figure~\ref{fig-samp-norm}.

\hypertarget{sec-app-fdr-ex}{%
\section{Projective contrasts for other summary statistics in
Oregon}\label{sec-app-fdr-ex}}

\begin{figure}

{\centering \includegraphics{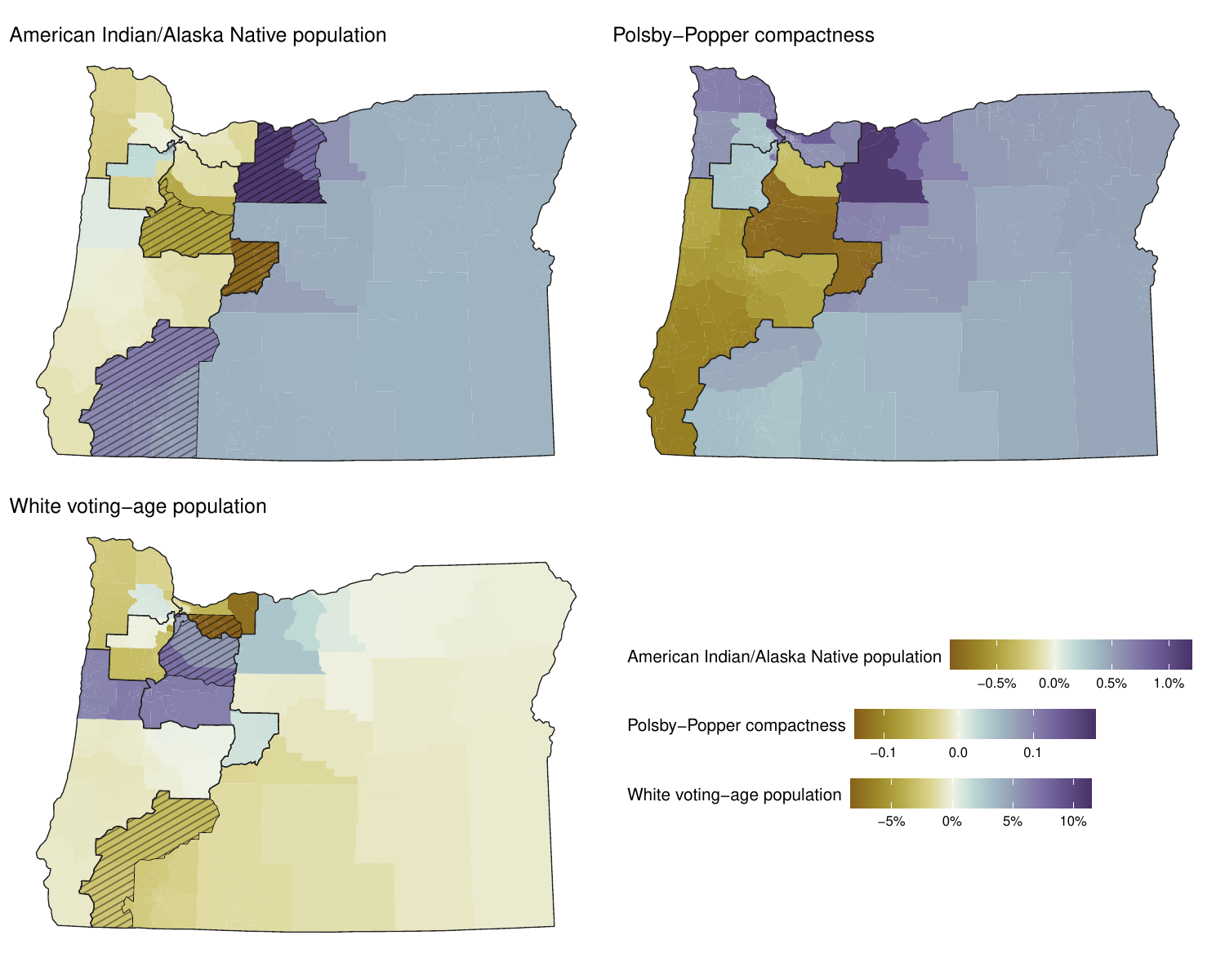}

}

\caption{\label{fig-fdr-addl-ex}Projective contrast of the enacted plan
in Oregon for three different summary statistics. Significant
discrepancies from the ensemble after controlling the pFDR at 5\% are
marked by the hatched areas. The enacted districts are indicated by the
black borders.}

\end{figure}

Figure~\ref{fig-fdr-addl-ex} shows the projective contrast maps for
three additional district-level summary statistics: the Polsby-Popper
compactness score, the American Indian / Alaska Native population, and
the White voting-age population. While the compactness projective
contrast initially appears to indicate that in many areas the enacted
plan is more or less compact than the typical ensemble plan, the pFDR
control procedure indicates that these differences are within the
expected range. In contrast, the racial composition of parts of
districts in the south, center, and north in the enacted plan are
significantly different from the composition of the corresponding areas
under the ensemble, as is clear from the pFDR-controlled projective
contrasts of AI/AN population and White VAP.

We stress that these results are only preliminary, and that
(pFDR-controlled) statistical significance, or lack thereof, does not
mean that the observed values of the projective contrast are large or
small in any absolute, substantive sense. Projective averages and
contrasts are a useful exploratory and interpretive tool regardless of
the number of areas selected by the ST procedure at a particular level
of pFDR control.

\end{document}